\documentclass{iau}
\usepackage{graphicx,subfigure}

\newcommand{\MJup}{$M_{\mathrm{Jup}}$}

\newcommand{\masyr}{$\mathrm{mas}\,\mathrm{yr}^{-1}$}

\title[The BASS Survey]
{The BANYAN All-Sky Survey for Brown Dwarf Members of Young Moving Groups}

\author[J. Gagn\'e, D. Lafreni\`re, R. Doyon, J. K. Faherty et al.]
{Jonathan Gagn\'e$^1$, David Lafreni\`ere$^1$, Ren\'e Doyon$^1$, Jacqueline K. Faherty$^{2,3,4}$, Lison Malo$^{6,1}$, Kelle L. Cruz$^{3,5}$, \'Etienne Artigau$^1$, Adam J. Burgasser$^7$, Marie-Eve Naud$^1$, Sandie Bouchard$^1$, John E. Gizis$^8$ \and Lo\"ic Albert$^1$}

\affiliation{
$^1$Institut de Recherche sur les Exoplan\`etes (iREx), Universit\'e de Montr\'eal, D\'epartement de Physique, C.P.~6128 Succ. Centre-ville, Montr\'eal, QC H3C~3J7, Canada.\\email: {\tt jonathan.gagne@astro.umontreal.ca}\\
$^2$Department of Terrestrial Magnetism, Carnegie Institution of Washington, DC~20015, USA\\
$^3$Department of Astrophysics, AMNH, Central Park West at 79th Street, New York, NY~10024\\
$^4$Hubble Fellow\\
$^5$Department of Physics \& Astronomy, Hunter College, City University of New York, 695 Park Avenue, NY~10065, USA\\
$^6$Canada-France-Hawaii Telescope, 65-1238 Mamalahoa Hwy, Kamuela, HI~96743, USA\\
$^7$CASS, UCSD, 9500 Gilman Dr., Mail Code 0424, La Jolla, CA~92093, USA\\
$^8$Dept. of Physics and Astronomy, Univ. of Delaware, 104 The Green, Newark, DE~19716, USA}

\pubyear{2015}
\volume{314}
\pagerange{119--126}
\jname{Young Stars \& Planets Near the Sun}
\editors{J. H. Kastner, B. Stelzer, \& S. A. Metchev, eds.}
\begin{document}

\maketitle

\begin{abstract}
We describe in this work the \emph{BASS} survey for brown dwarfs in young moving groups of the solar neighborhood, and summarize the results that it generated. These include the discovery of the 2MASS~J01033563--5515561~(AB)b and 2MASS~J02192210--3925225~B young companions near the deuterium-burning limit as well as 44 new low-mass stars and 69 new brown dwarfs with a spectroscopically confirmed low gravity. Among those, $\sim$\,20 have estimated masses within the planetary regime, one is a new L4\,$\gamma$ bona fide member of AB~Doradus, three are TW Hydrae candidates with later spectral types (L1--L4) than all of its previously known members and six are among the first contenders to low-gravity $\geq$\,L5\,$\beta$/$\gamma$ brown dwarfs, reminiscent of WISEP~J004701.06+680352.1, PSO~J318.5338--22.8603 and VHS~J125601.92--125723.9~b. Finally, we describe a future version of this survey, \emph{BASS-Ultracool}, that will specifically target $\geq$\,L5 candidate members of young moving groups. First experimentations in designing the survey have already led to the discovery of a new T dwarf bona fide member of AB~Doradus, as well as the serendipitous discoveries of an L9 subdwarf and an L5 + T5 brown dwarf binary.
\end{abstract}

\firstsection

\section{Introduction}

Young Moving Groups (YMGs) are ideal laboratories to study the low-mass end of the initial mass function and obtain age-calibrated samples of brown dwarfs to understand the evolution of their fundamental and atmospheric properties over time. However, the identification of substellar members of YMGs is challenging because of their intrinsic faintness and sparse distribution on the sky. These YMGs include e.g. the TW~Hydrae association (TWA; 5--15\,Myr; \cite[de la Reza et al. 1989]{1989ApJ...343L..61D}; \cite[Kastner et al. 1997]{1997Sci...277...67K}), Tucana-Horologium (THA; 20--40\,Myr; \cite[Torres et al. 2000]{2000AJ....120.1410T}) and AB~Doradus (ABDMG; 110--130\,Myr; \cite[Zuckerman et al. 2004]{2004ApJ...613L..65Z}).

\cite[Malo et al. (2013)]{2013ApJ...762...88M} developed the Bayesian Analysis for Nearby Young AssociatioNs (BANYAN) tool to identify new low-mass star members of YMGs in the solar neighborhood. The biggest power of this tool is that its use of Bayes' theorem allows it to derive membership probabilities even in scenarios where not all kinematic information is available. In particular, radial velocity and distance measurements are not available for most low-mass stars, which is primarily due to their relative faintness.

Our team built on this tool to create BANYAN~II (\cite[Gagn\'e et al. 2014a]{2014ApJ...783..121G}), an updated version of the tool that is tailored for the search of brown dwarfs, improved its inherent models of the field and YMGs and made it more efficient so that it became applicable to large data sets. This allowed us to identify 24 new candidate members of YMGs as well as a new M7.5 bona fide member of Tucana-Horologium among known low-gravity brown dwarfs.

\section{The \emph{BASS} Survey}

\begin{figure}[t]
	\center
	\subfigure[XYZ]{\includegraphics[width=0.495\textwidth]{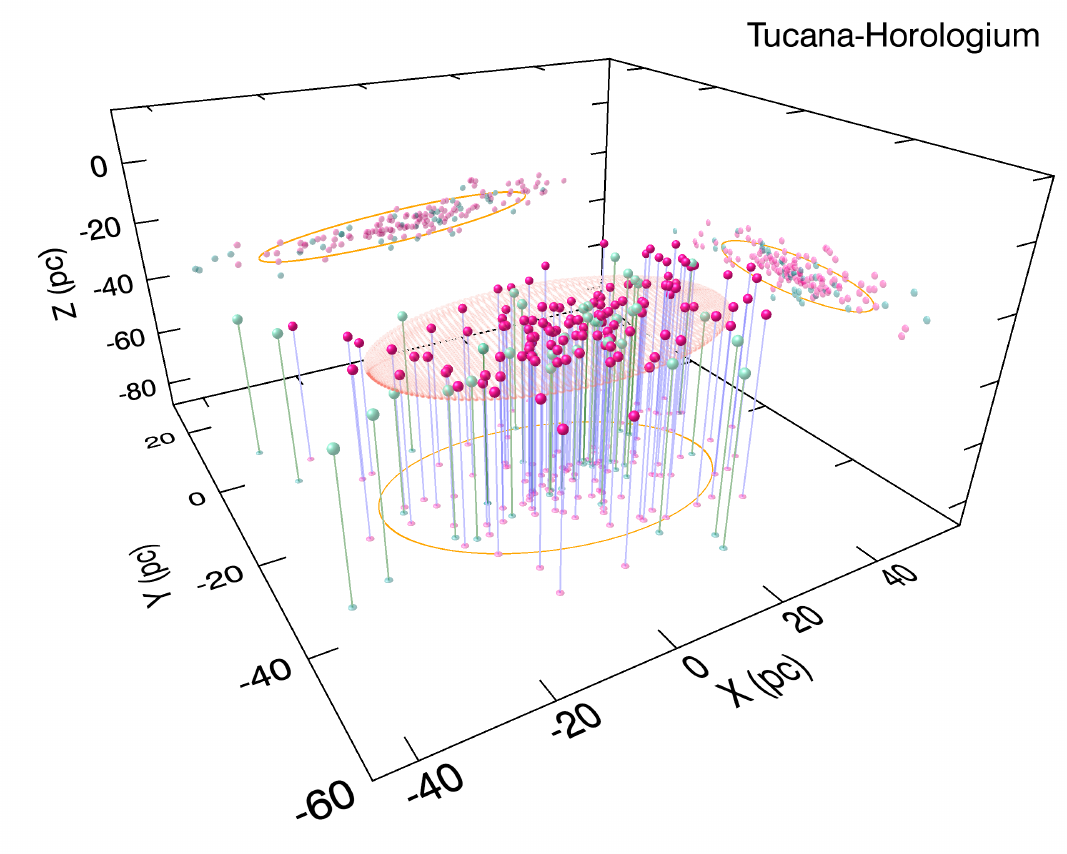}}
	\subfigure[UVW]{\includegraphics[width=0.495\textwidth]{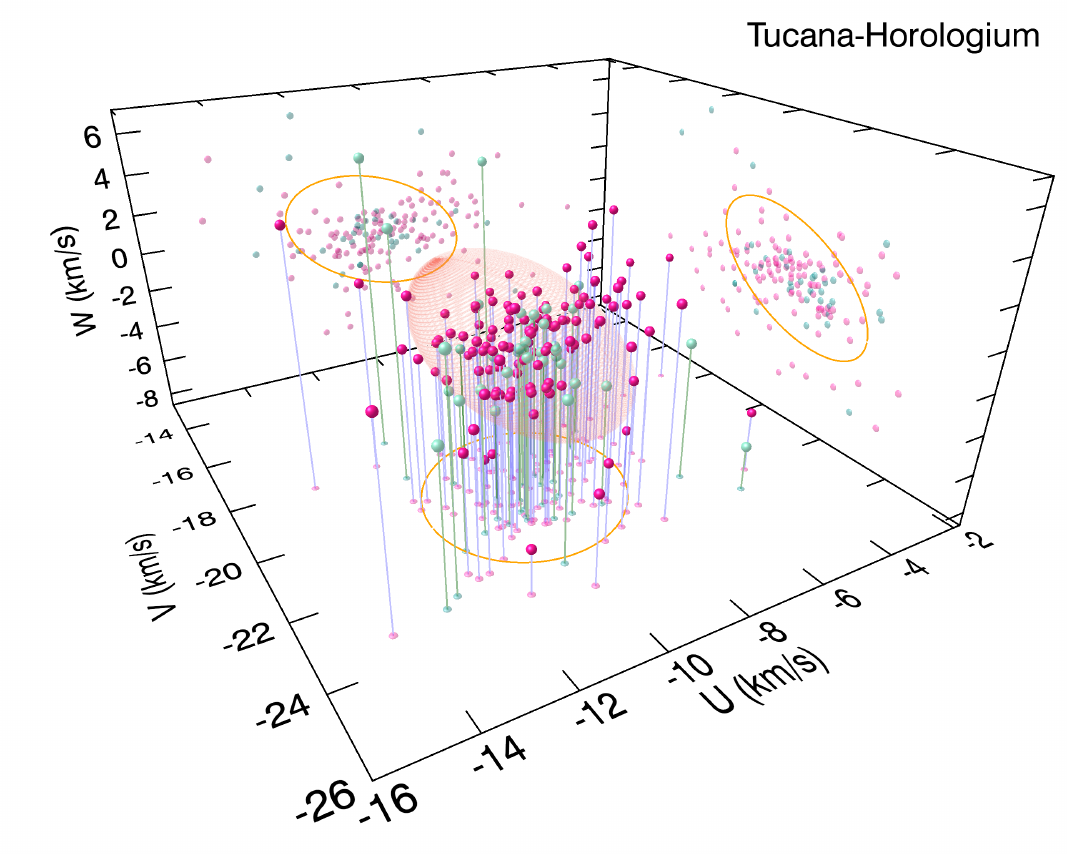}}
	\caption{Most probable galactic positions XYZ and space velocities UVW based on BANYAN~II statistical distances and RVs for all \emph{BASS} candidate members of THA (red points), compared with bona fide members (green points), as well as the spatial and kinematic ellipsoid models used in BANYAN~II (orange ellipsoids; see \cite[Gagn\'e et al. 2014a]{2014ApJ...783..121G}). All points and models are projected on the three normal planes for a better clarity.}
	\label{fig:xyzuvw_abdmg}
\end{figure}
\begin{figure}[t]
	\center
	\subfigure[New Low-Gravity Brown Dwarfs]{\includegraphics[width=0.54\textwidth]{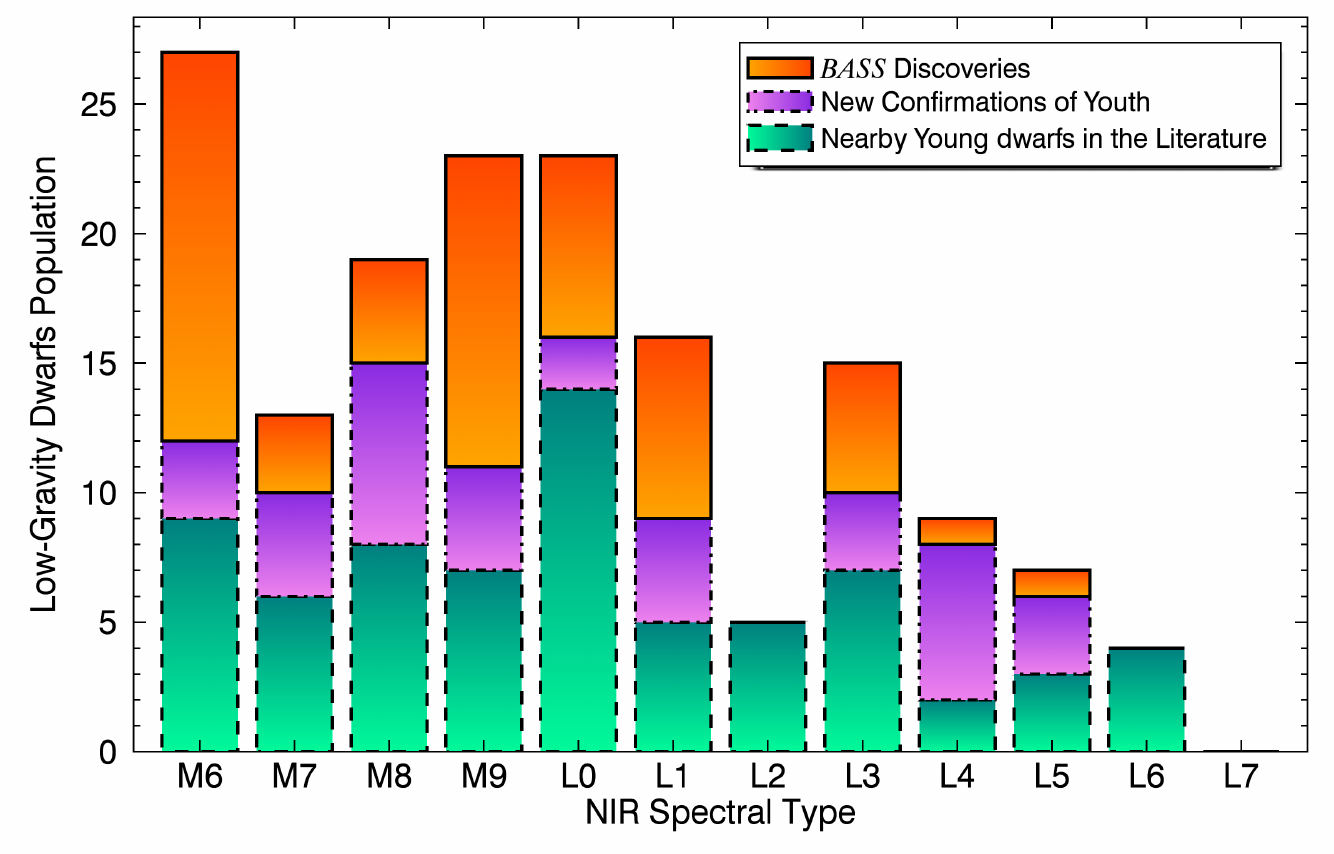}\label{fig:spts}}
	\subfigure[A new L4\,$\gamma$ member of ABDMG]{\includegraphics[width=0.45\textwidth]{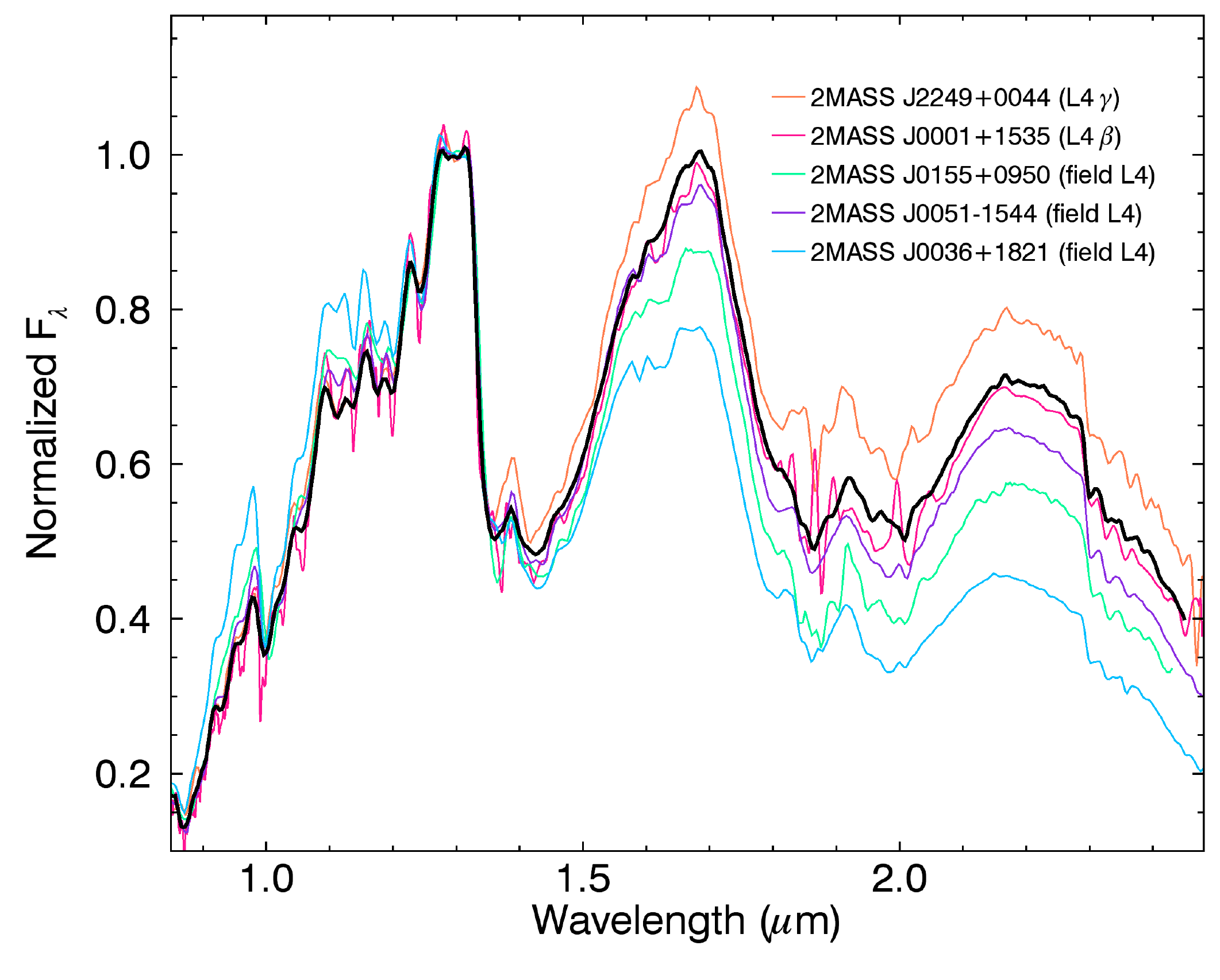}\label{fig:L4ABDOR}}
	\caption{Panel~a~: Near-infrared spectral types for all known low-gravity dwarfs and those presented in this work. The \emph{BASS} survey has contributed significantly in increasing the number of known low-gravity M6--L5 dwarfs. Panel~b~: Near-infrared spectrum of a L4\,$\gamma$ ABDMG bona fide member, compared with various field and low-gravity L4 brown dwarfs. Its $H$-band continuum has a triangular shape and a red continuum, which are signs of a low gravity.}
\end{figure}

Equipped with the BANYAN~II tool, we have undertaken the \emph{BANYAN All-Sky Survey} (\emph{BASS}), a survey for new brown dwarf candidate members of YMGs from a cross-match of the \emph{Two Micron All-Sky Survey} (\emph{2MASS}; \cite[Skrutskie et al. 2006]{2006AJ....131.1163S}) and the \emph{AllWISE} survey (\cite[Kirkpatrick et al. 2014]{2014ApJ...783..122K}) outside of the galactic plane. We used this cross-match to calculate proper motions with a typical precision of $\sim$\,15\,\masyr\ and to built an input list of 98\,970 high-proper motion objects with colors that are consistent with $\geq$\,M5 dwarfs in the solar neighborhood. We then used the BANYAN~II tool to identify 228 new late-type candidate members of YMGs, including 79 potential brown dwarfs and 22 potential isolated planetary-mass objects. A cross-match with the literature allowed us to estimate our rate of false positives at around $\sim$\,13\% in this survey. An additional list of 275 candidates (the \emph{Low-Priority BASS}, or \emph{LP-BASS}, sample) was generated with the same method except with more permissive color cuts.

In Figure~\ref{fig:xyzuvw_abdmg}, we display the most probable galactic positions (XYZ) and space velocities (UVW) for all candidate members of THA that we identified in the \emph{BASS} survey. The BANYAN~II tool outputs the most likely statistical distances and radial velocities for each target, which allowed us to derive their most probable XYZ and UVW. It can be noted that our candidate members are likely located in the same loci as bona fide members of THA. In the XYZ plane, they also follow the plane-shaped distribution of THA members, a property of THA members that was first observed by \cite[Kraus et al. (2014)]{2014AJ....147..146K}.

As part of this survey, we have identified a low-gravity M6 low-mass star candidate member of THA which was identified as a binary star with a 12--14\,\MJup\ companion at a separation of 84\,AU (\cite[Delorme et al. 2013]{2013A&A...553L...5D}), as well as an additional low-gravity M6\,$\gamma$ + L4\,$\gamma$ system (with a separation of 160\,AU) also in THA \cite[(Artigau et al. 2015)]{2015arXiv150501747A}. We also identified the latest-type L dwarf candidate member of TWA (\cite[Gagn\'e et al. 2014b]{2014ApJ...785L..14G}) and a $\sim$\,10\,\MJup\ low-gravity mid-L candidate member of Argus (\cite[Gagn\'e et al. 2014c]{2014ApJ...792L..17G}) in the first stage of the near-infrared spectroscopic follow-up of the \emph{BASS} sample. More recently, we presented the spectroscopic follow-up of a significant portion of \emph{BASS} (J. Gagn\'e et al., submitted to ApJS), revealing 44 new low-mass stars and 69 brown dwarfs with spectroscopic confirmations of low gravity (Figure~\ref{fig:spts}). A new L4\,$\gamma$ bona fide member of AB~Doradus was also discovered as part of this work (see Figure~\ref{fig:L4ABDOR}).

\begin{figure}
	\centering
	\subfigure[Distribution of Estimated Masses]{\includegraphics[width=0.495\textwidth]{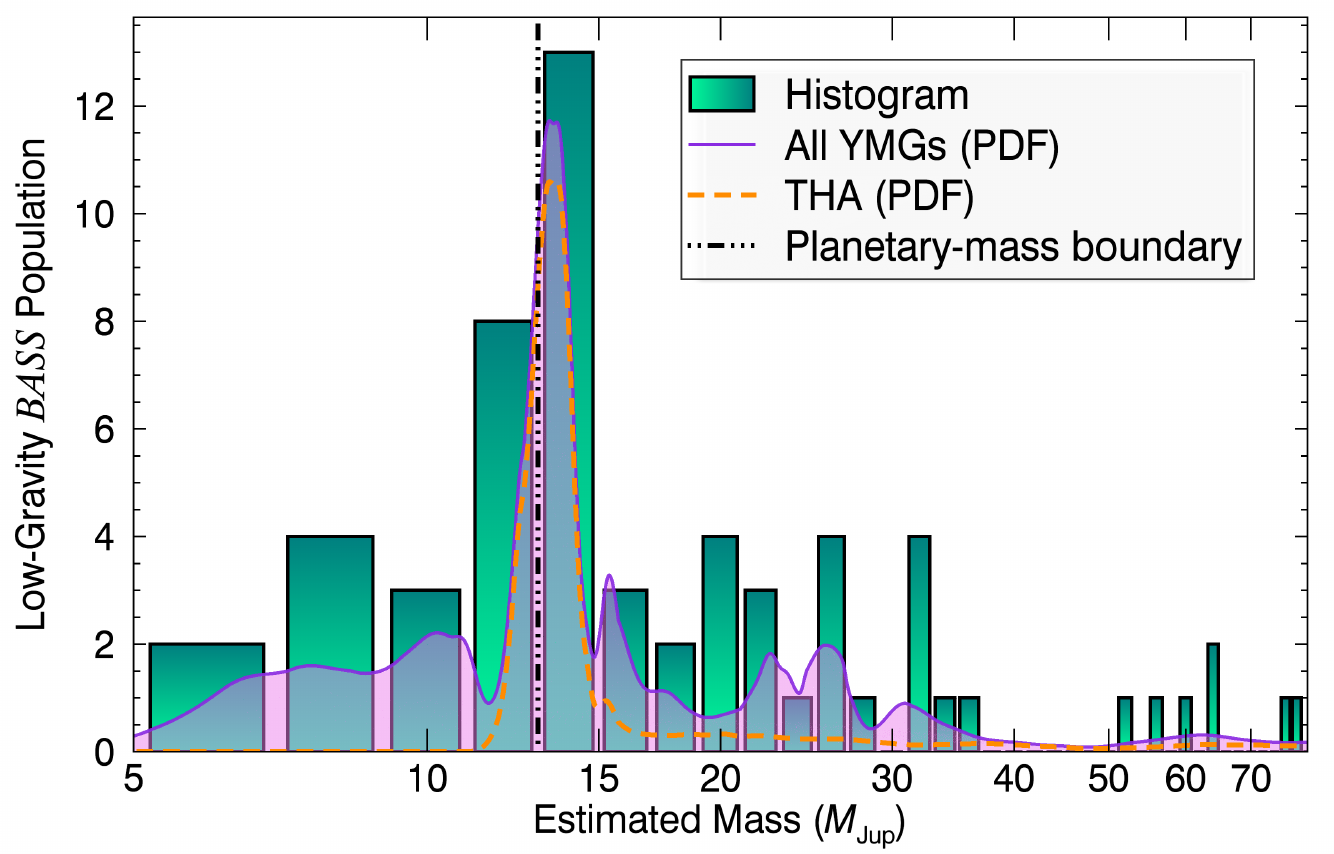}\label{fig:histmass}}
	\subfigure[Color-Magnitude Diagram]{\includegraphics[width=0.45\textwidth]{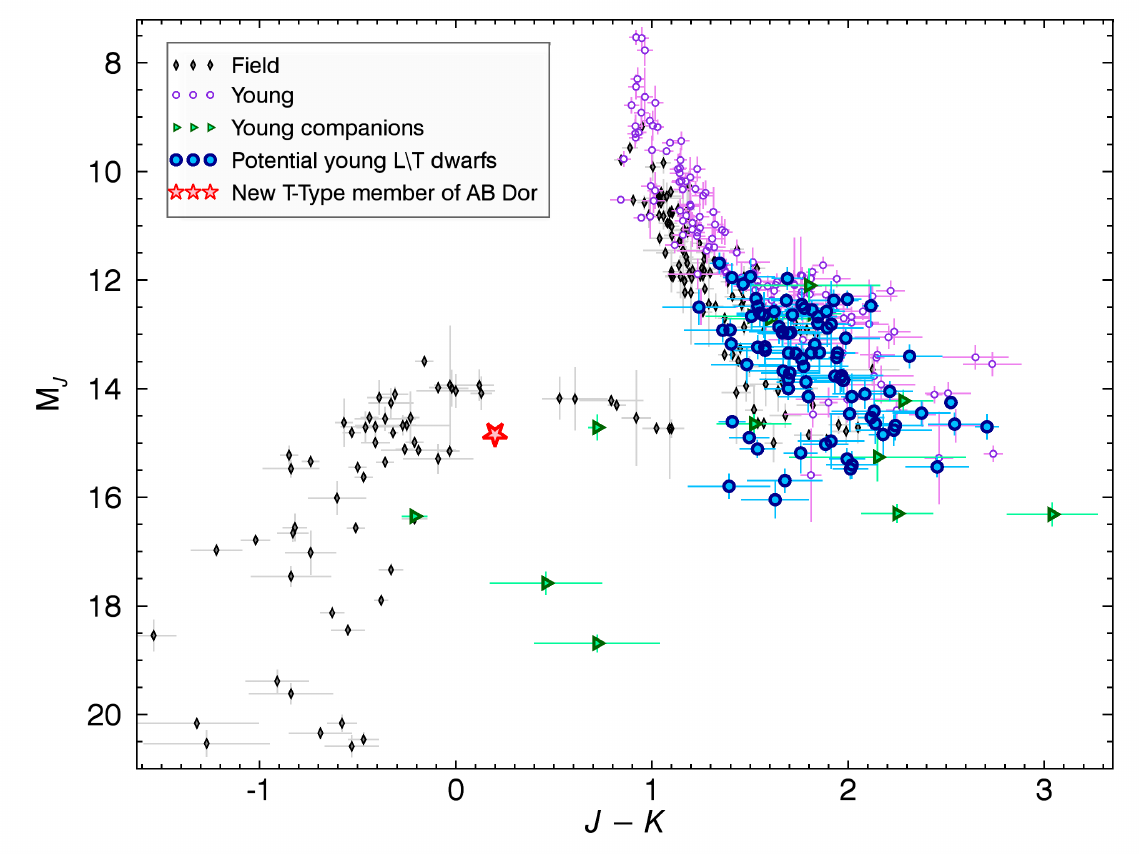}\label{fig:cmd}}
	\caption{Panel a~: Estimated masses for low-gravity dwarfs in our sample. The continuous probability density functions (PDFs) provide a histogram-like continuous distribution that include measurement errors and are independent of binning. Panel~b~: MKO $J-K$ versus $M_J$ color-magnitude diagram for field and low-gravity brown dwarfs. A few known substellar companions, new YMG candidates at the L/T transition from a preliminary version of \emph{BASS-Ultracool} and a new bona fide T dwarf member of ABDMG are also shown.}
\end{figure}

We used these new discoveries and known low-gravity brown dwarfs to create new color-magnitude diagrams (e.g., Figure~\ref{fig:cmd}) and investigate their fundamental properties using the BT-Settl atmosphere models (\cite[Allard et al. 2013]{2013MSAIS..24..128A}). We identified an unexpectedly large number of 12--14.5\,\MJup\ spectroscopically confirmed low-gravity candidate members of THA, where our current follow-up is most sensitive, compared with its main-sequence stellar population (Figure~\ref{fig:histmass}). This represents an over-density of a factor $36.4_{-12.5}^{+16.6}$. We find that systematic errors in the atmosphere models are not likely to cause such an over-density, and that the most likely explanation would be a combination of small number statistics and young interlopers from YMGs other than THA. It will be necessary to obtain radial velocity and parallax measurements to assess whether this is the case or if THA really has an up-turn in the low-mass end of its IMF.

We are currently working on an updated version of BANYAN~II that will include additional young associations with slightly older ages (up to $\sim$\,500\,Myr) and combine this with a cross-match of \emph{2MASS}, \emph{AllWISE}, \emph{SDSS} and \emph{UKIDSS} to identify new $\geq$\,L5 young brown dwarfs, up into the T dwarfs regime. Our first tests in constructing this survey have already allowed us to identify a new T dwarf bona fide member in AB~Doradus. It is shown in a color-magnitude diagram in Figure~\ref{fig:cmd} and this result will be presented in J. Gagn\'e et al. (submitted to ApJL).

\section{Conclusions}

We reviewed in this work some discoveries from the BANYAN All-Sky Survey that aims at the identification of new low-gravity brown dwarf members of moving groups in the solar neighborhood, and outlined future plans to locate low-gravity T dwarfs in the upcoming years. These discoveries will serve as important benchmarks to understand the atmospheric properties of giant, gaseous exoplanets and to characterize the low-mass end of the initial mass function.

\bibliographystyle{apj}

\clearpage
\begin{discussion}

\discuss{Katelyn Allers}{It seems that you have a lack of low-gravity L2 dwarfs (Figure 2). What explains this ?}

\discuss{Author}{This is partly due to our lack of an L2\,$\beta$ template and the fact that we assign spectral types by comparison to visual templates. We have some low-gravity L2 contenders, but they have a low signal-to-noise ratio, so we refrained from calling them L2 for the moment until we have a good template (i.e. those would be called L1: or L3: right now). This "valley" at L2 becomes much less apparent if we include spectral type uncertainties to generate a continuous histogram (a probability density function) and the difference between the number of L1/L2 or L2/L3 low-gravity objects becomes insignificant at 0.2$\sigma$.}

\discuss{Mark Pinsonneault}{On the over-density of low-mass candidates in Tucana-Horologium; you are exploring a region of the color-magnitude diagram that can be populated by mergers and weird objects that might mimick the properties of the objects you are really searching for.}

\discuss{Author}{I agree that there are such things in that part of the color-magnitude diagram, but I would be really surprised if such a variety of objects could mimick the full spectra of young brown dwarfs at resolutions of $R \sim$\,100-6000. I would believe that for these peculiar contaminants, the strength of absorption lines would be different, that there would be some strong emission lines or that we would see some reddening. Obtaining radial velocities and parallaxes will definitely tell us without a doubt if we are observing such objects, and this is something that I want to do next.}

\discuss{Gilles Chabrier}{I think that those numbers [over-density of low-mass candidates] are not very significant because of small number statistics.} 

\discuss{Author}{We used Poisson statistics to estimate the error bars on our measurements and on the over-density factor; it seems that it is a 3$\sigma$ result at the moment, however combining this with interlopers from other moving groups could really bring down the significance. For example if only 5/12 of our candidates turn out to be members of other young associations then the over-density would become just a 1$\sigma$ result. Our simulations to characterize the false-positive rates in BANYAN~II tell us that it's almost impossible that these kinds of candidates are interlopers (all have $>$\,90\% membership in Tucana-Horologium and are confirmed as younger than $\sim$\,200\,Myr), however this assumes that our models of the neighborhood (i.e. the field and its moving groups) are perfect, and this is definitely not the case. We could be missing some groups, and some known moving groups might be shaped differently than what we think, which could thus very well cause a fraction of these 12 objects to be young interlopers.}

\end{discussion}

\end{document}